# The practice of self-citations: a longitudinal study

Silvio Peroni, silvio.peroni@unibo.it, https://orcid.org/0000-0003-0530-4305
Research Centre for Open Scholarly Metadata & Digital Humanities Advanced Research Centre (DHARC), Department of Classical Philology and Italian Studies, University of Bologna, Bologna, Italy

Paolo Ciancarini, paolo.ciancarini@unibo.it, https://orcid.org/0000-0002-7958-9924
Department of Computer Science and Engineering, University of Bologna, Bologna, Italy and Innopolis University, Innopolis, Russia

Aldo Gangemi, aldo.gangemi@unibo.it, https://orcid.org/0000-0001-5568-2684
Digital Humanities Advanced Research Centre (DHARC), Department of Classical Philology and Italian Studies, University of Bologna, Bologna, Italy

Andrea Giovanni Nuzzolese, andrea.nuzzolese@istc.cnr.it, https://orcid.org/0000-0003-2928-9496
Semantic Technology Laboratory (STLab), Institute of Cognitive Science and Technologies, National Research Council, Rome, Italy

Francesco Poggi, francesco.poggi5@unibo.it, https://orcid.org/0000-0001-6577-5606
Department of Computer Science and Engineering, University of Bologna, Bologna, Italy

Valentina Presutti, valentina.presutti@cnr.it, https://orcid.org/0000-0002-9380-5160
Semantic Technology Laboratory (STLab), Institute of Cognitive Science and Technologies, National Research Council, Rome, Italy

**Corresponding author:** Silvio Peroni, silvio.peroni@unibo.it, +39 051 20 9 8576, via Zamboni 32, 40126 Bologna (BO), Italy

**Abstract**
In this article, we discuss the outcomes of an experiment where we analysed whether and to what extent the introduction, in 2012, of the new research assessment exercise in Italy (a.k.a. Italian Scientific Habilitation) affected self-citation behaviours in the Italian research community. The Italian Scientific Habilitation attests to the scientific maturity of researchers and in Italy, as in many other countries, is a requirement for accessing to a professorship. To this end, we obtained from ScienceDirect 35,673 articles published from 1957 and 2016 by the participants to the 2012 Italian Scientific Habilitation, that resulted in the extraction of 1,379,050 citations retrieved through Semantic Publishing technologies. Our analysis showed an overall increment in author self-citations (i.e. where the citing article and the cited article share at least one author) in several of the 24 academic disciplines considered. However, we depicted a stronger causal relation between such increment and the rules introduced by the 2012 Italian Scientific Habilitation in 10 out of 24 disciplines analysed.

**Article Highlights**
- ~6.5% of the citations were author self-citations, where the citing article shared at least one author with the cited article
- We found that 21 out of 24 disciplines used to group the articles analysed had an increment in self-citations after 2012
- Ten disciplines showed a pronounced increment of self-citations that could be caused by the rules introduced in the 2012 Italian Scientific Habilitation

**Acknowledgements**
We want to thank our colleagues of the Digital and Semantic Publishing Laboratory at the University of Bologna for their support and discussions on the topic – namely Angelo Di Iorio and Fabio Vitali. Also, we would like to extend our thanks to Marzia Freo and Alessandra Luati (University of Bologna) for providing specific statistical backgrounds and techniques we used in the analysis presented in this paper. Last but not least, Andrea Bonaccorsi (University of Pisa) provided essential insights on the project and Riccardo Fini (University of Bologna) made available to us its SQL database containing information about the people who participated in the 2012 Italian Scientific Habilitation. ANVUR has partially funded this work.

# Introduction

Citations are the *conceptual directional links* from citing entities to cited entities (i.e. *PaperA cites PaperB*), which are instantiated by the inclusion of a bibliographic reference in the reference list of the citing entity or in one of its footnotes (Peroni & Shotton 2018a). Nowadays, citations are still the dominant measurable unity of credit in science (Fortunato et al. 2018). They are one of the main clues used by researchers for gaining knowledge about a particular topic. They are used by scientists in Bibliometrics, Informetrics, and Scientometrics to analyse the complex relationships that exist within the scientific landscape. They also serve institutional goals since they provide one of the primary mechanisms for assessing the quality of research through impact metrics and indicators calculated from citation databases.

However, despite the very general definition just mentioned, citations are not born all equal. For instance, each citation instantiated in a particular in-text reference pointer (e.g. "[4]" or "(Fortunato et al. 2018)") carries a particular *function*. Such function is the reason why a particular paper is cited (to extend its outcomes, to criticise it, to refer to it as related work, etc.) (Teufel, Siddharthan & Tidhar 2006), as shown in Fig. 1. Furthermore, the kinds of citation characterisations are identifiable according to other factors. Citations in which the citing and the cited entities have something relevant in common with one another, over and beyond their subject matter, for example, authors, journal, institutional affiliation, or funding agency, are named *self-citations*. In the past, several works – e.g. (Aksnes 2003) (Glänzel & Thijs 2004) (Costas, van Leeuwen & Bordons 2010) (Bartneck & Kokkelmans 2011) (Di Iorio et al. 2015) (Larivière et al. 2015) (Gálvez 2017) (Gul, Shah & Shafiq 2017) – have provided discussions, insights, analyses and tools about this kind of citations in different contexts.

A concise review of some of the primary studies concerning self-citations is presented in (Glänzel et al. 2006)[1]. In particular, this study presents the role of self-citations from two different perspectives. On the one hand, when we describe citations as a measure of *reception* of scientific results, self-citations must be considered as part of the citation behaviour of the authors within the scholarly domain, since they describe, even, the way authors' ideas and research have evolved in time. On the other hand, when we see citations as a proxy for expressing the impact or even the quality of a research work that may result in some *reward* for the authors, self-citations indeed may distort and, potentially, even falsifying the impact of particular research, scholar or journal. As some examples, see the simulation proposed by Bartneck and Kokkelmans (2011) analysing the strategic creation of self-citations for incrementing the *h-index*, and the 2013 case of *citation stacking* in some Brazilian journals (Van Noorden 2013).

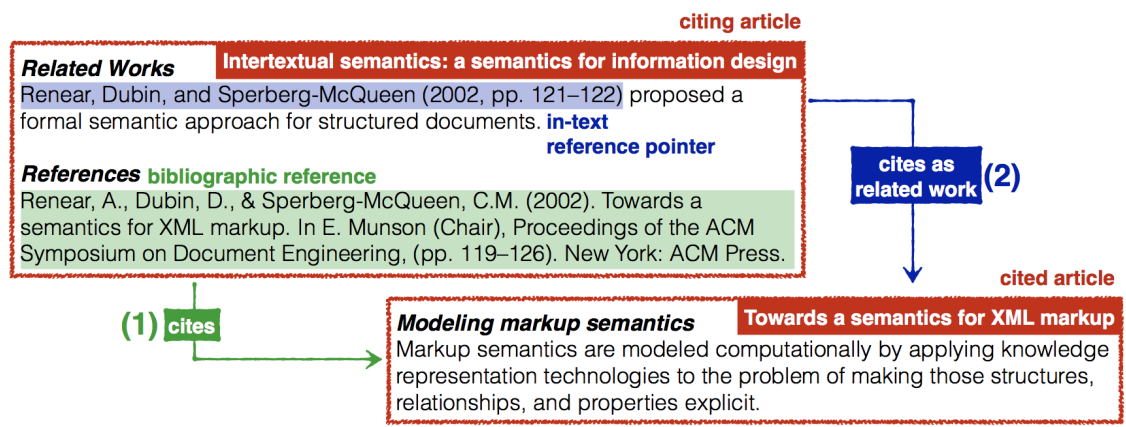

*Fig. 1* The figure shows two possible kinds of citation links. A bibliographic reference is responsible for the instantiation of one citation link (1), which expresses the purely syntactic connection between a citing article and a cited article. Instead, an in-text reference pointer (denoting a bibliographic reference) instantiates the other citation link (2), which is characterised by the function the citation is conveying – i.e. the reason why the author cites a particular article.

In our study, the perspective adopted for self-citations is the latter of the ones as mentioned earlier, i.e. the one related with the reward of authors, since we are explicitly referring to the outcomes of the Italian research assessment exercise called Italian Scientific Habilitation. In particular, two kinds of self-citations are of interest for our work: *author self-*

---

[1] It is worth mentioning that (Glänzel et al. 2006), as well as other cited studies on self-citations, define the term *citations* of a particular article as the number of other articles that cite it, by including it in their reference list. On the contrary, as introduced at the beginning of this section, we use the term *citations* for indicating the *links* between a citing entity and a cited one, as defined in (Peroni & Shotton 2018a). In the former case, we say that the definition of term *citations* is *article-centric* since it strictly depends on the particular article one is considering, while in the latter case we say that the definition is *relational-centric* since it cares only about the connection between two entities.

*citations* and *author network self-citations*². As shown in Fig. 2, citations of the former type are those where the citing article and the cited article share at least one author. Citations of the latter type occur when a co-author of any author of the citing article is also the author of the cited article.

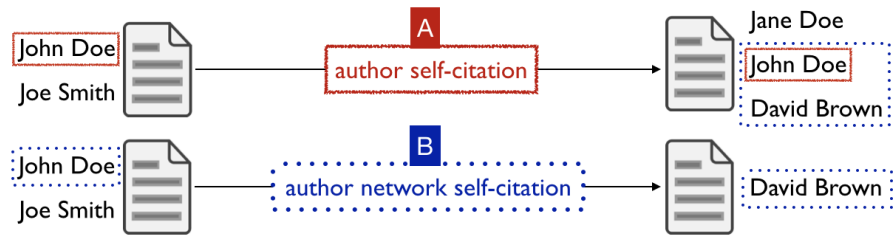

**Fig. 2** *The figure shows two particular kinds of self-citations. Citation A introduces an author self-citation, i.e. where the citing article and the cited article share at least one author: John Doe is the author of both the citing and cited articles in the figure. Citation B introduces an author network self-citation, where a co-author of any of the authors of the citing article is also the author of the cited article. In this case, John Doe co-authored an article (the cited one of citation A) with David Brown, and the latter one also authored the cited article in citation B.*

In this work, we want to understand if the rules used in the Italian Scientific Habilitation exercises have an impact on the use of the self-citations shown in Fig.2. As a concrete example, we have analysed the citation behaviours of the scholars who participated in the 2012 Italian Scientific Habilitation exercise.

As a consequence of the latest revision of the Italian university law, only people who have attained the Italian Scientific Habilitation can apply for tenured positions in Italian universities. Nevertheless, obtaining the Italian Scientific Habilitation does not guarantee any position by itself.

The first two sessions of the Italian Scientific Habilitation took place in 2012 and 2013 and involved more than 50,000 candidates. The rules introduced in 2012 by ANVUR (the Italian National Agency for the Evaluation of the University and Research Systems) for the Italian Scientific Habilitation do not exclude author self-citations and author network self-citations for the computation of all the metrics for assessing researchers participating to the Habilitation. The goal of this work is to investigate what has been the impact of such rules on self-citations habits considering the people participating in the 2012 Italian Scientific Habilitation. In particular, the goal of our study is to answer the following research questions (RQs from now on):

- RQ1: Did the Italian Scientific Habilitation programme affect the number of author self-citations? Moreover, if so, in which disciplines such change has been more pronounced?
- RQ2: Did the Italian Scientific Habilitation programme affect the number of author network self-citations? Moreover, if so, in which disciplines such change has been more pronounced?

The setting of our work, i.e. the Italian Scientific Habilitation, has been already discussed and analysed in the past – e.g. see (Marzolla 2015, 2016) (Seeber et al. 2018) (Baccini, De Nicolao & Petrovich 2019) (Di Iorio, Peroni & Poggi 2019) (Nuzzolese et al. 2019) (Poggi et al. 2019). In particular, Baccini, De Nicolao and Petrovich (2019) showed that the recent increase of self-citations recorded at the national level, looking at the affiliations of the scholars, seemed to be a peculiar trend in Italy. The novelty in our analysis is to consider author self-citations and author network self-citations on a broad set of scientific disciplines (24 in total, according to the classification provided by Scimago Journal Rank) by looking at the articles authored by the participants to the Italian Scientific Habilitation. In particular, we gathered citation information defined in the reference lists of all the articles available in XML on ScienceDirect within a 1957-2016 publication window that were authored by any of the participants to the 2012 Italian Scientific Habilitation exercise. We employed novel technologies related to the Semantic Publishing domain (Shotton 2009) for gathering and analysing these citation data.

## The Italian Scientific Habilitation: how it works

In 2011, the Italian Law 240/2010 introduced extensive changes in which the university system, and in particular its academics, are evaluated. The main change has been the replacement of the position of Assistant Professor with two other fixed-term positions called Junior Assistant Professor and Senior Assistant Professor (in Italian, *Ricercatore a Tempo*

---

² The definition of these and other kinds of self-citations has been taken from CiTO, the Citation Typing Ontology available at http://purl.org/spar/cito, part of the SPAR Ontologies (Peroni & Shotton 2018b). We used (Wallace, Larivière & Gingras 2012) and from the blog post "Journal self-citations are increasingly biased toward impact factor years" by Ludo Waltman and Caspar Chorus (https://www.cwts.nl/blog?article=n-q2x264) to derive the definitions of some of the self-citations described in such ontology.

*Determinato di Tipo A* e *Ricercatore a Tempo Determinato di Tipo B* respectively). The junior position lasts three years, plus two optional additional years. The senior position is a proper tenure track – it lasts three years, and then, if the candidate is compliant with particular requirements, he/she will become Associate Professor. In particular, in order to be eligible to become an Associate Professor, the scholar must have been obtained the Italian Scientific Habilitation (in Italian, *Abilitazione Scientifica Nazionale*, or *ASN*), i.e. a certificate which attests the scientific maturity of the scholar for getting to that specific position.

A potential candidate can participate to two distinct Habilitation procedures, i.e. the one that assesses if he/she has the requirements to become an Associate Professor and the one that assesses the requirements to become a Full Professor. It is worth mentioning that obtaining the ASN does not guarantee to get a tenure position, but having it is a mandatory requirement for accessing to a professorship.

Besides, the academic organisation in Italy considers 14 distinct Scientific Areas (i.e. academic disciplines, such as Mathematical Sciences and Informatics, Physical Sciences, Political and Social Sciences). Each Scientific Areas can, in turn, includes one or more Recruitment Fields (in Italian, *Settore Concorsuale*, such as Computer Science, Astronomy, Sociology). Each of these Recruitment Fields corresponds to a particular academic field of study. Each candidate can apply for the Habilitation to one or more Recruitment Fields at the same time. Since each field has its own rules for assessing its candidates, the candidate may get the Habilitation only for some of his/her applications – usually, just for one, i.e. his/her "natural" one.

Furthermore, such academic organisation distinguishes between *bibliometric* and *non-bibliometric* Recruitment Fields. In fact, on the one hand, part of the evaluation of some of these Recruitment Fields (e.g. Computer Science) is based on bibliometrics indicators, in particular, (a) the total number of journal articles, (b) the number of citations received, and (c) the h-index normalised according to the scientific age of the candidate. Two of these indicators – i.e. (b) and (c) – are firmly based on citations received by the products authored by the scholars. On the other hand, other Recruitment Fields (e.g. any field in the Arts and Humanities) use different kinds of indicators for evaluating the scholars in disciplines and, in particular, do not use citations as the primary metric.

## Methods and Material

The initial bibliographic and citation data we used in our analysis have been gathered and analysed through Semantic Publishing technologies. Semantic Publishing (Shotton 2009) concerns the use of Web and Semantic Web technologies and standards for enhancing a scholarly work (e.g. using plain RDF statements (Cyganiak, Wood & Lanthaler 2014)) to improve its discoverability, interactivity, openness and (re-)usability for both humans and machines. The assumptions of openness implicit in Semantic Publishing have been explicitly adopted for the publication of research data by the FAIR (Findable, Accessible, Interoperable, Re-usable) data principles (Wilkinson et al. 2016). Early examples of the semantic enrichment of scholarly works involved the use of manual (e.g. see (Shotton et al. 2009)) or (semi-)automatic post-publication processes (e.g. see (Peroni 2017)).

Semantic Publishing technologies allow one to enrich the semantic payload of the networks of published articles, usually linked through their plain citation links, in order to describe several content-related and context-related aspects of the publishing domain. For instance, it would be possible to group such semantic enrichment in eight different buckets. Each bucket can be able to describe a particular semantic specification of an article. For instance, it can concern either the description of the article content from different angles (e.g. structure, rhetoric, argumentation) or contextual elements relating to the creation of a paper (e.g. research project, people contributions, publication venue) (Peroni 2017).

Recently, Semantic Publishing technologies have been used in bibliometric and scientometrics studies – for instance, see (Huang et al. 2019) (Nuzzolese et al. 2019) (Poggi et al. 2019) (Xiao, Shi, Z. & Wang 2018) (Zhu et al. 2019). Also, several infrastructures dedicated to the publication of a vast amount of bibliographic and citation data – such as OpenCitations (Peroni & Shotton 2020) and WikiData (Vrandečić & Krötzsch 2014) – has started to use Semantic Publishing technologies for making available, queryable, interlinkable, and interoperable their data.

### Semantic Publishing technologies used

In this work, we have adopted specific Semantic Publishing technologies for retrieving the data needed for our analysis. In particular, we have modified two tools that implement the ingestion workflow of the OpenCitations Corpus (Peroni, Shotton & Vitali 2017) – which is an open repository of RDF-based scholarly citation data: the *Bibliographic Entries Extractor* (BEE) and the *SPAR Citation Indexer* (SPACIN).

BEE was responsible for retrieving the article metadata and the references of all the articles available on ScienceDirect in XML written by the participants of the 2012 Italian Scientific Habilitation. Overall, we obtained 35,673 XML documents. It is worth noticing that the availability of these articles in XML was subject to the particular limits and constraints related to a contract between the University of Bologna and Elsevier (owner of ScienceDirect). The choice of considering only articles that were available with their full-text XML for the analysis was twofold. On the one hand, this work has been run in the context of a research project partially funded by ANVUR in 2015, with the idea of being open to possible follow-up projects using the data collected for the analysis. One of the future works, discussed in "Conclusions", concerns the processing and analysis of the context where a self-citation is done in the text, to retrieve the implicit citation function (Teufel, Siddharthan & Tidhar 2006) related to that citation that we can annotate on the XML source using specific technologies we have developed for this purpose such as EARMARK (Barabucci et al. 2013). This analysis would enable us to study whether self-citations are used to convey an organic or perfunctory function (Swales 1986) according to the sections of the articles containing them. On the other hand, in addition to the possibility of having XML documents available for processing, we chose to use ScienceDirect (i.e. Elsevier) publications only to have a good and representative population of articles across several distinct disciplines – guaranteed by the massive presence of Elsevier as one of the leading publishers, in terms of journal publications, in all the disciplines analysed.

SPACIN was responsible for retrieving additional metadata information about all the retrieved citing/cited articles by querying the Crossref API (for additional metadata) and the ORCiD API (for additional information about the authors of the article). These APIs were also used to disambiguate bibliographic resources and agents using the identifiers retrieved (e.g., DOI, ISSN, ISBN, ORCiD, URL, and Crossref member URL) or by using existing services available through such APIs. In particular, we entirely relied on the Crossref textual query service to retrieve a bibliographic entity (and its metadata) from a pure textual bibliographic reference included in an article. Also, we relied on the ORCiD API to retrieve all the ORCiDs of the authors of an article identified by a DOI. It is worth mentioning that, due to the nature of the services used for retrieving and enhancing the bibliographic and citation data, only the articles involved in DOI-to-DOI citations had a full set of metadata specified, that were functional to our analysis. Thus, it would be possible that some citations between articles were not considered in the study. We used a local triplestore, i.e. a database to store RDF data that makes them queryable via SPARQL (Harris & Seaborne 2013), to store all these metadata, once retrieved.

These technologies made it very easy to perform additions and analysis on the metadata retrieved, through their flexibility in manipulating and extending data. In particular, we added to all the citing articles additional information about the main disciplines related to all the journals published by Elsevier, by using publisher's official list of active journals and their subject categories (https://www.elsevier.com/__data/promis_misc/sd-content/journals/jnlactivesubject.htm). We used the Scimago Journal Rank (http://www.scimagojr.com/) to retrieve missing information about the journals dismissed or that have changed their name in the past.

## Kinds of self-citations considered

All the citations we obtained by using SPACIN are *synchronous* (Aksnes 2003), i.e. those an article gives using the bibliographic references in its reference list. Indeed, all the citations that we retrieved were extracted from the reference lists of the articles available in ScienceDirect written by the participants of the Italian Scientific Habilitation. Thus, every citing article was written by at least one participant to the 2012 Italian Scientific Habilitation.

We decided to use synchronous citations since it allowed us to observe possible changes in the behaviour of self-citations around the years when the new rules of the Italian Scientific Habilitation were introduced (i.e. 2012). In particular, we decided to split the citations into two populations according to the date of publications of the articles we retrieved: all the citations defined in the articles published by the end of 2012, and those defined in the article published after 2012. We marked up all these citations as *author self-citations* and *author network self-citations* according to the following rules.

The marker "author self-citation" was applied if there was at least one overlap between a family name of an author of the citing article and a family name of an author of the cited article. We used family names for identifying author self-citations according to the following intuition. When an author with a particular family name cites, in one of its articles, another article in which one of the authors share the same family name, it is highly probable that they are referring to the same person. Intuitively speaking, authors tend to cite themselves, while citations to homonymous are very unlikely. This situation seems to occur in our dataset, which deals with the Italian domain only that is characterised by a vast plethora of different family names. We have empirically tested this intuition by retrieving 100 random citations which share, in the citing article and the cited article, at least one author's family name. In all the randomly selected citations, all the authors with the same family name referred to the same person[3]. For the sake of clarity, the family names used for marking a citation as an author self-citation may refer to authors who could have not participated to the 2012 Italian Scientific Habilitation.

---

[3] We decided not to use the ORCiD data retrieved for testing the precision and recall of the heuristics based on family names for identifying author self-citations due to the partial coverage of the ORCiDs assigned to the authors of the articles considered in the analysis. To build a robust gold set using ORCiD, we needed that all the authors in all the citing and cited articles in such set had an ORCiD specified. However, this is not the case of the ORCiD dump we have used for the analysis, where we did not find some citing and cited articles with all the authors having the ORCiDs assigned. In addition, we were aware that the family name approach for identifying author self-citations can be unreliable for people coming from particular countries, such as Asian authors. Even if this situation did not happen in the empirical test we run, it was possible in principle and, as a consequence, could have slightly distorted some (even if a limited set of) self-citation counts.

We were also interested in the author network self-citations independently from the year of publication of the citing article and the cited articles they involved. However, in this case, we cannot follow the same intuition adopted for the author self-citations to identify them. In fact, since creating the co-author cluster for a particular article using the family names of the involved authors would have resulted in biased data, creating a cluster much larger than it actually would be. Considering the full name (i.e. given name plus family name) for disambiguating the authors was not an optimal solution either since a lot of given names were recorded only with the initials. This situation would have resulted in colliding authors with similar names as the same person. Thus, the solution we used for reducing the possible false positives for these kinds of citation links is to use the ORCiD identifiers of the authors as the primary (and precise) mechanism to assess if the authors of two articles involved in a citation referred to the same person. However, we did not retrieve a reasonable amount of ORCiD identifiers for having complete coverage due to the lack of available public data in the ORCiD repository.

Finally, it is worth mentioning that our study considered self-citations at the article level, by marking a citation between two articles as a self-citation when such articles shared a particular feature (either an author or an author network, as illustrated above). In particular, we did not adopt an author-centric perspective (i.e. by counting self-citations performed by individual authors). Possible limitations derived from this choice are introduced in Section "Discussion".

## Citation data

By using the data mentioned above, we analysed the citations to the references contained in 35,673 articles published from 1957 to mid-2016, organised in 24 disciplines – where each article was associated with at least one discipline. Overall, we obtained 1,379,050 citations (~39 citations per article). We used two different conditions in our analysis.

The first one considered the number of author self-citations (cf. RQ1 in Section "Introduction") and author network self-citations (cf. RQ2 in Section "Introduction") identified in the articles published from 1957 to 2012 (26,951 articles) against those ones identified in the articles published from 2013 to 2016 (8,722 articles). The second condition considered the self-citations in the article published in 2013-2016 against a similar time frame, i.e. the self-citations of the articles published in 2009-2012 (8,511 articles).

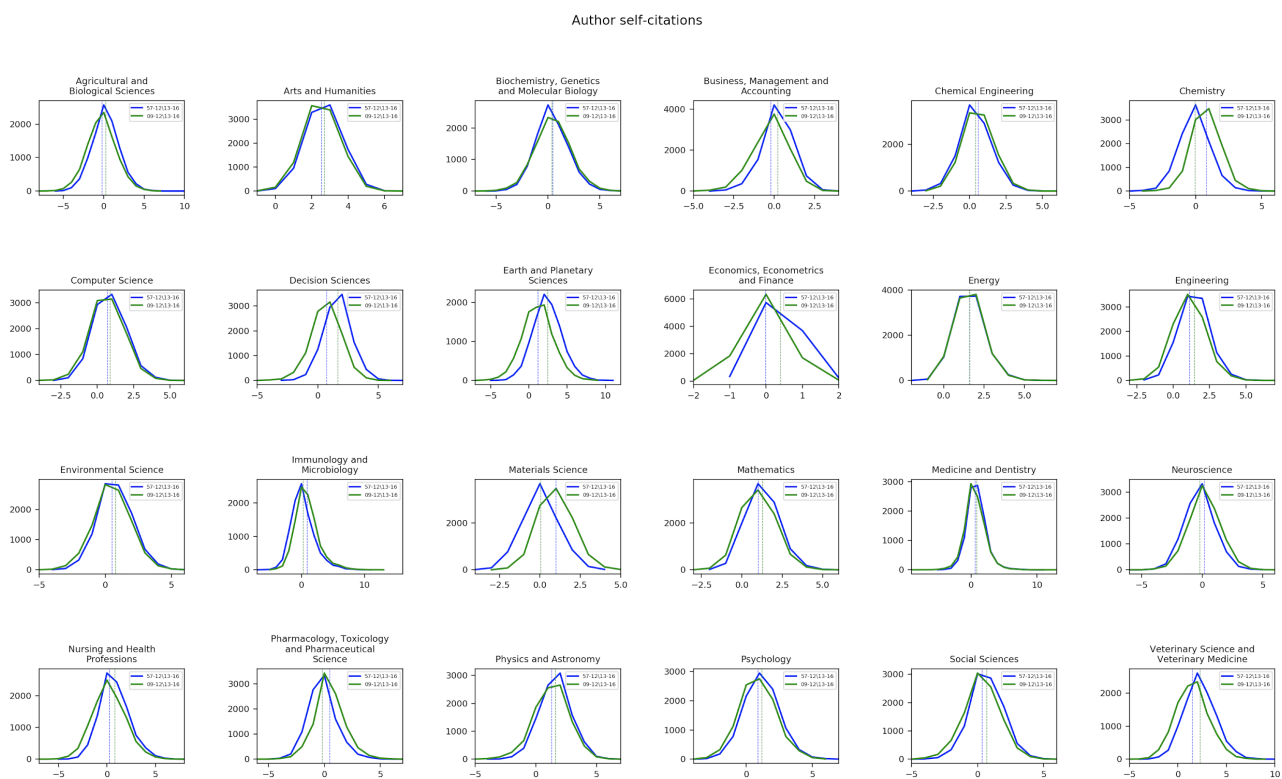

*Fig. 3* The diagrams, for each discipline, showing the differences of the means of author self-citations (x-axis) considering the two distinct conditions of our analysis, i.e. the articles published in 1957-2012 against those ones published in 2013-2016 (in blue), and those ones published in 2009-2012 against the ones published in 2013-2016 (in green). The figures were calculated empirically by creating 10,000 distinct samples from the self-citation data available in our dataset (y-axis).

For each of these populations, we calculated the mean of author self-citations and author-network self-citations per article and the related standard deviation. Also, we added to such statistics the difference of the means of the self-citations between the populations accompanied by the related 95% confidence intervals. It is worth mentioning that, while the distribution of the author self-citations and author-network self-citations of each discipline was lognormal, the distribution of the differences of the means followed a normal distribution. We proved the existence of such distributions by creating 10,000 distinct samples from the self-citation data available in our dataset, that were drawn on a diagram to show this behaviour. For instance, Fig. 3 shows the diagrams generated for the author self-citations. We observed similar behaviour in the diagram showing the author network self-citations.

### Linear regression on self-citations

We also calculated the linear rate of change of self-citations up to 2012, by using a linear regression approach which considered the average number of self-citations per year. Such linear rate of change was useful to understand if increments in self-citations after 2012 had some relations with the rules introduced in the 2012 Italian Scientific Habilitation or derived from global changes (e.g. the adoption of a "publish or perish" mentality worldwide). In particular, we predicted the expected rate of self-citation up to 2016, and we compared such expectation with the actual observed rate of self-citation in the 2013-2016 time frame and with the related linear rate of change of self-citations computed by considering the data of such latter four years.

It is worth mentioning that the data used as input of the linear regression algorithm adopted[4] could contain outliers (i.e. years with very high/low average of self-citations). However, we did not remove them from the data used for calculating the linear rate of change to avoid the introduction of possible undesired bias.

## Results

In this section, we discuss the results related to two aspects of the data gathered, i.e. author self-citations and author network self-citations, according to the two distinct conditions, i.e. the comparison using the articles published in 1957-2012 and 2013-2016, and those published in 2009-2012 and 2013-2016. All the data, as well as the software developed for creating the figures presented in this section, are available in (Peroni 2018).

### Analysis of author self-citations

Considering the overall citations obtained (1,379,050 citations) in all the articles published from 1957 to 2016, we found that 91,398 were author self-citations (~6.5% of the total). The mean of self-citations per article of each population – i.e. 1957-2012 and 2013-2016 – was calculated, and the difference of such means has shown an increasing number of self-citations after 2012, except in some specific disciplines: *Neuroscience*, *Chemistry*, and *Pharmacology, Toxicology and Pharmaceutical Science*. We accompanied such mean with the standard deviation and confidence intervals, as shown in Table 1 (in Appendix).

In order to understand if the increment of author self-citations was present also considering a stricter publication time-window, we have also analysed the citations defined by the references contained in 17,233 articles published from 2009 to 2016, still organised in 24 disciplines and split into two populations – i.e. 2009-2012 and 2013-2016. We obtained 759,217 citations (~44 citations per article), of which 48,586 were self-citations (~6.4% of the total) – similar to that one measured for all the articles in our dataset. Table 2 (in Appendix) shows all the statistics about these two populations. Again, we observed an increment of the number of self-citations after 2012, except in some specific disciplines: *Agricultural and Biological Sciences*, *Business, Management and Accounting*, and *Economics, Econometrics and Finance* – which are different from those highlighted when considering the full article dataset from 1957 to 2016.

In Fig. 4, the difference of the means of author self-citations for each discipline is shown, accompanied by the related confidence interval, highlighted using the error bars. We used two vertical dashed lines for indicating the value where no difference is measured (the red dashed line on the left of each diagram) and the mean difference value considering all the disciplines (the blue dashed line on the right of each diagram). According to the 1957-2016 period, we found that 21 out of 24 disciplines had an increment in self-citations after 2012, with 15 of these having the confidence interval greater than 0, where we observed a clear increment of self-citations. We had similar results also analysing the disciplines in the smallest period (2009-2016), where we had that 21 out of 24 disciplines had an increment in self-citations, and 17 of them having a confidence interval which did not overlap with the "no difference" threshold.

In order to try to understand if such increments, highlighted in Fig. 4, were related to the rules introduced in the 2012 Italian Scientific Habilitation or to other global factors (e.g. the worldwide "publish or perish" behaviour), we calculated the expected author self-citation rate in the 2013-2016 period as introduced in Section "Methods and Material". We compared such expected author self-citation rate with the actual author self-citation rate observed in the 2013-2016 period.

---

[4] We used the Python class `LinearRegression` defined in the `sklearn.linear_model` package of *scikit-learn* (https://scikit-learn.org).

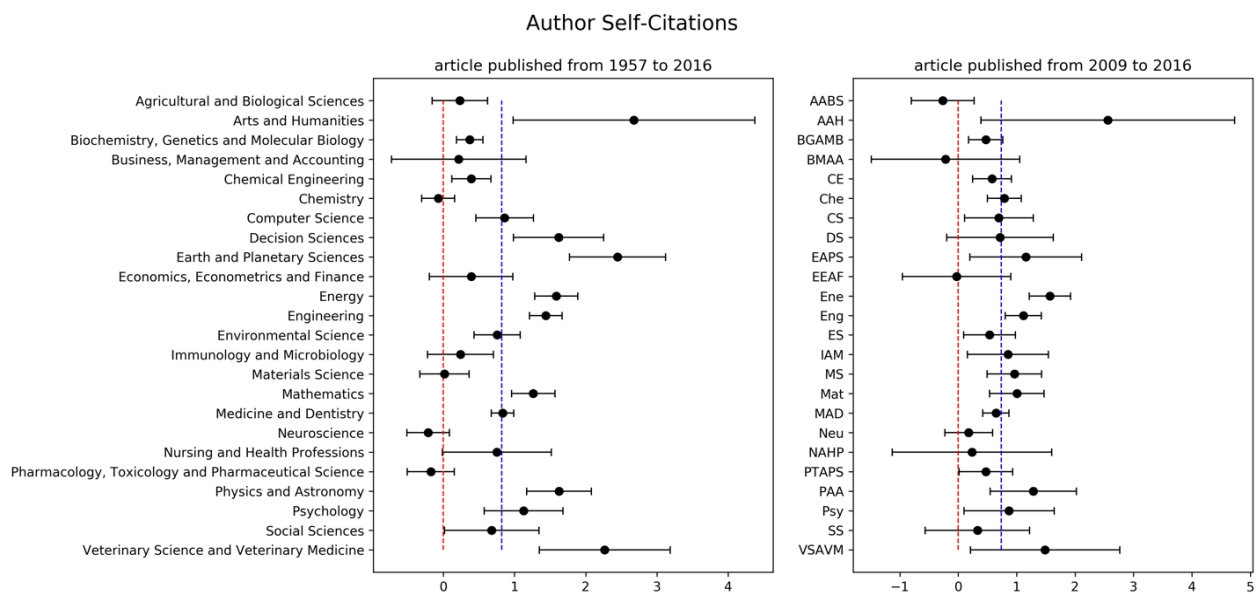

**Fig. 4** Confidence intervals of the difference of the means of author self-citations of the considered populations for each discipline, taking into consideration the articles published from 1957 to 2016 and from 2009 to 2016. The two populations used to compute such difference included, respectively, all the articles published by 2012 and all the articles published since 2013. The vertical red dashed line indicates where there was no difference in the means, while the vertical blue dashed line shows the mean of all the differences between all the disciplines. In the diagram on the right, the name of the disciplines are recalled by using their acronym obtained by using the first three letters for single word disciplines (e.g. "Eng" for "Engineering") and the first letter of each word in the related compound name of a discipline (e.g. "CS" for "Computer Science").

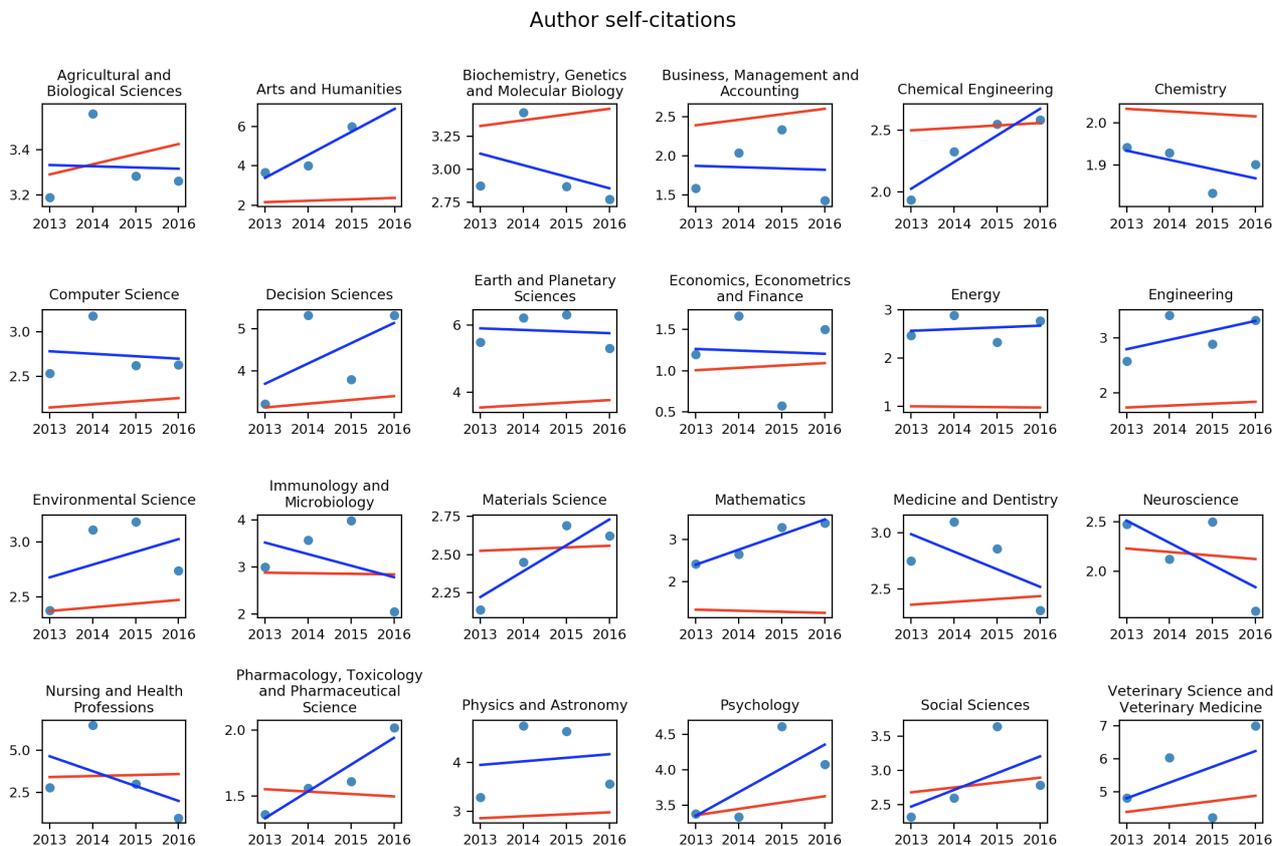

**Fig. 5** The figure shows the expected author self-citation rate in the 2013-2016 period predicted from the self-citation data available from 1957 to 2012 (red line), and the actual author self-citation rate observed in the 2013-2016 (light blue dots) and its related linear regression trend (blue line).

In the period 1957-2016 (shown in Fig. 5), only 11 of the 15 cases showing a clear increment of self-citations (i.e. those having the related confidence interval greater than 0) also shown a clear increment of self-citations compared with the expected self-citation citation rate after 2012. In particular, in 9 of these (e.g. *Arts and Humanities* and *Mathematics*), the increment registered in the 2013-2016 period was even highlighting a more increasing trend against the one calculated using the self-citations available since 2012. Instead, for 2 of them (i.e. *Chemical Engineering* and *Social Science*), the actual increment compared with the expected rate started after a few years. Finally, 2 of the 9 cases (according to Fig. 4), where we observed either a decrement of self-citations or a difference of means with the confidence intervals overlapping with 0 (i.e. Materials Science and Pharmacology, Toxicology and Pharmaceutical Sciences), showed a clear increasing trend in author self-citations since 2015.

In the period 2009-2016 (shown in Fig. 6) the number of cases having a clear increment of self-citations according to the expected citation rate after 2012 is lesser than the one observed in the 1957-2016 period – i.e. 9 out of 17. Eight of these (e.g. *Arts and Humanities* and *Mathematics*) recorded a clear increasing trend, while in another one (i.e. *Veterinary Science and Veterinary Medicine*) the expected and actual trends for the self-citations in the 2013-2016 window were pretty close. However, they seemed to register an inversion in 2016, were the actual trend seemed to surpass the expected one. Finally, 4 of the 7 cases in which the confidence intervals (shown in Fig. 3) overlapped with 0 (e.g. *Social Science*s and *Economics, Econometrics and Finance*) shown a clear increasing trend in author self-citations between 2013 and 2014.

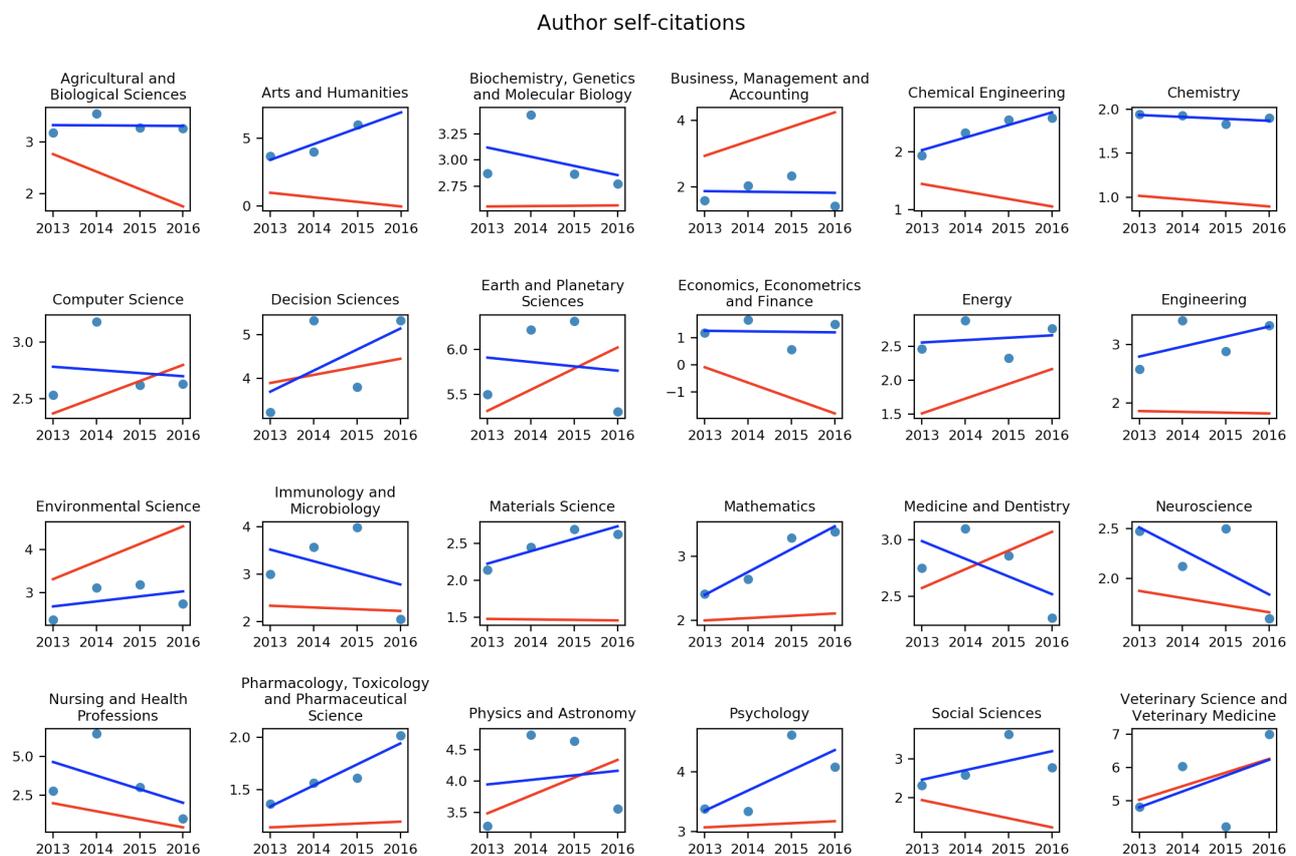

***Fig. 6*** *The figure shows the expected author self-citation rate in the 2013-2016 period predicted from the self-citation data available from 2009 to 2012 (red line), and the actual author self-citation rate observed in the 2013-2016 (light blue dots) and its related linear regression trend (blue line).*

## Analysis of author network self-citations

Considering the overall citations obtained (1,379,050 citations) in all the articles published from 1957 to 2016, we found that 21,317 were author network self-citations (~1.5% of the total). The mean of self-citations per article of each population – i.e. 1957-2012 and 2013-2016 – was calculated, and the difference of such means has shown a slightly moderate increasing number of author network self-citations overall, as shown in Table 3 (in Appendix).

Similarly, we also analysed the citations defined by the references contained in 17,233 articles published from 2009 to 2016, and we obtained 13,242 author network self-citations (~1.17% of the total). In this case, the difference of means of such self-citations per article between the two populations – i.e. 2009-2012 and 2013-2016 – was close to 0, as shown in Table 4 (in Appendix).

In Fig. 7, the difference of the means of author network self-citations for each discipline is shown, accompanied by the related confidence interval, highlighted using the error bars. Also, we have computed the expected author network self-citation rate in the 2013-2016 period, predicted from the self-citation data available since 2012. We compared it with the actual author network self-citation rate observed in the 2013-2016 period, but we did not obtain any relevant situation – and, thus, we avoid to report here the related graphs.

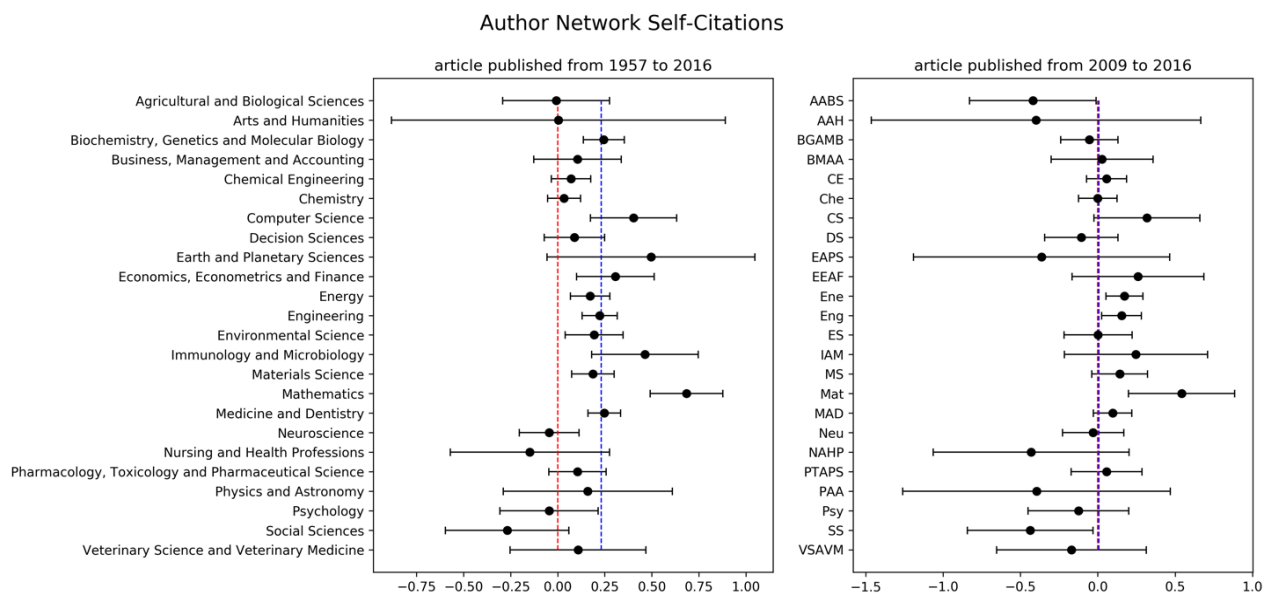

***Fig. 7*** *Confidence intervals of the difference of the means of author network self-citations of the considered populations for each discipline, taking into consideration the articles published from 1957 to 2016 and from 2009 to 2016. The two populations used to compute such difference included, respectively, all the articles published by 2012 and all the articles published since 2013. The vertical red dashed line shows where there was no difference, while the vertical blue dashed line shows the mean of all the differences between all the disciplines. In the diagram on the right, the name of the disciplines are recalled by using their acronym obtained by using the first three letters for single word disciplines (e.g. "Eng" for "Engineering") and the first letter of each word in the related compound name of a discipline (e.g. "CS" for "Computer Science").*

## Discussion

In this section, we discuss the results introduced in the previous sections, focussing on the outcomes obtained for the author self-citations. For what concern the author network self-citations, the data retrieved and the observations presented in the previous section seemed to highlight that we were not able to gather enough information for an accurate analysis. This situation was mainly due to the lack of precise and complete information for identifying the authors (i.e. the small amount of ORCiD retrieved), which affected the correct identification of the co-author network of a given author. We plan to explore this aspect further in future studies by exploiting additional open datasets (e.g. Wikidata).

### Publications between 1957 and 2016

Considering only the disciplines in which the confidence interval of the values in Fig. 4 was not overlapping with 0 (15 disciplines out of 24), we observed an apparent incremental behaviour from 0.36 additional author self-citations (in *Biochemistry, Genetics and Molecular Biology*) to 2.68 additional author self-citations (in *Arts and Humanities*) recorded after 2012. The disciplines that had the most substantial increment in self-citations per article were *Arts and Humanities* (from 1.66 to 4.33), *Earth and Planetary Science* (from 3.45 to 5.9), and *Veterinary Science and Veterinary Medicine* (from 3.21 to 5.48).

Even if these three disciplines were those that had recorded the most substantial increment in terms of author self-citations, their confidence interval shows a quite large variability of the real potential value of the related difference between the means of self-citations. This behaviour can be due to the limited number of articles published after 2012 we had available for these disciplines – this aspect is particularly evident in the *Arts and Humanities* discipline. However, according to Fig. 5, we noticed a decreasing trend in the author self-citations between 2013-2016, when compared with the expected one, only in the *Earth and Planetary Science* discipline. We noticed similar behaviour in three other disciplines – i.e. *Biochemistry, Genetics, and Molecular Biology*, *Computer Science*, and *Medicine and Dentistry*. This situation may suggest that the increment recorded in these disciplines was not so far from the expected self-citation value for such disciplines.

The disciplines *Arts and Humanities* and *Social Science* (where we observed an increment of 0.68 author self-citations per article after 2012, confirmed by the trend analysis shown in Fig. 5) are interesting for another aspect. The evaluation guidelines introduced in The Italian Scientific Habilitation commended not to use citation metrics for assessing the quality of the articles of these two disciplines, but rather other quantitative and qualitative factors. However, it seemed that the fact that the Italian Scientific Habilitation considered author self-citations in "bibliometric" disciplines such as Mathematics had implicitly brought the authors of the articles in Arts and Humanities and Social Sciences to increment the number of self-citations. Thus, this behaviour seemed to happen although the Italian Scientific Habilitation committees of non-bibliometric disciplines did not use citations for assessing the candidates.

Instead, *Energy*, *Engineering*, *Mathematics* and *Physics and Astronomy* were among the disciplines showing a difference between the means of the two populations higher than the average (i.e. 0.82). Besides, these disciplines also showed very smooth confidence intervals that do not overlap with the average mentioned above and a clear increasing trend after 2012, as shown in Fig. 5.

The data gathered also shown that, overall, there is an increasing trend of self-citations in time – as highlighted in the red regression lines in Fig. 5 calculated considering the self-citations found in the articles published from 1957 to 2012. However, we recorded a slightly small decreasing trend in self-citations in six disciplines: *Chemistry*, *Energy*, *Immunology and Microbiology*, *Mathematics*, *Neuroscience*, and *Pharmacology, Toxicology and Pharmaceutical Science*. Past events may have influenced this behaviour, e.g. whether some assessment rules of past evaluation exercises conveyed an increment of author self-citations or even a decrement in case the impact of an article was somehow balanced considering the amount of self-citation it contained.

## Publications between 2009 and 2016

Even considering the shorter period, between 2009 and 2016, the trend in author self-citations is still significantly increasing. 17 disciplines out of 24 (as shown in Fig. 4) had the confidence interval that was not overlapping with 0. In particular, we recorded an average increment of 0.74 author self-citations per article and, among the 17 disciplines mentioned above, the dimension of such increment went from 0.47 in *Biochemistry, Genetics and Molecular Biology* to 2.56 author self-citations per an article in *Arts and Humanities*. Those that showed the most substantial increment were *Arts and Humanities* (from 1.77 to 4.33), *Energy* (from 1.04 to 2.61), and *Veterinary Science and Veterinary Medicine* (from 3.99 to 6.17).

Only two disciplines, i.e. *Energy* and *Engineering* (that registered an increment from 1.9 to 3.01 author self-citations per article), had (small) confidence intervals that surpassed, without any overlapping, the mean difference between all the disciplines (i.e. the blue line in Fig. 4). However, according to the trend analysis shown in Fig. 6, *Energy* showed a quite controversial behaviour since the regression line (in red) calculated considering the self-citations until 2012 seemed to cross the line (in blue) derived from the actual self-citation data from 2013 to 2016. We observed similar behaviour in other seven disciplines – i.e. *Biochemistry, Genetics and Molecular Biology*, *Computer Science*, *Earth and Planetary Science*, *Environmental Science*, *Immunology and Microbiology*, *Medicine and Dentistry*, and *Physics and Astronomy*. Thus, according to this, it may be possible that such increment in author self-citations measured in this short period on the disciplines mentioned above could depend on factors unrelated with the rules introduced in the 2012 Italian Scientific Habilitation. Alternatively, it is even possible that such disciplines reacted to the new rules introduced by the 2012 Italian Scientific Habilitation even before 2012 since they were (even informally) known a few years before their actual use in 2012.

Among all the disciplines, there is one – i.e. *Agricultural and Biological Sciences* – that showed a quite clear decreasing trend, with a confidence interval that slightly overlaps with 0. In this case, the number of self-citations before and after 2012 decreased by 0.26 author self-citations per article. However, looking at the self-citation trends in Fig. 6, the trend of the author self-citations within the 2013-2016 period calculated for this discipline (blue line) is very different from the one expected (red line). This behaviour could suggest that in one of the past four years, i.e. within the 2009-2012 period, the discipline recorded a quite high number of self-citations, even if it is not clear if the rules introduced by the new Italian Scientific Habilitation concerned with this situation.

## Cross-comparison

Comparing the outcomes of author self-citations in the 1957-2016 period against the 2009-2016 period, we noticed as the latter one had a higher number of disciplines (17 against 15, as in Fig. 4) that shown a confidence interval that was not overlapping with 0. At first sight, this could be supportive of the hypothesis that the 2012 Italian Scientific Habilitation rules had indeed affected the number of self-citations done after 2012. However, the comparison of the expected vs actual trends in author self-citations (as in Fig. 5 and Fig. 6) are not so favourable in the 2009-2016 period, since we recorded controversial trends in 8 disciplines against the 4 cases highlighted in the 1957-2016 period. This scenario could be a sign that, even if we observed an increment, the citation data gathered in the shorter period were not enough for supporting the behavioural change in self-citations actively.

The disciplines (*Agricultural and Biological Sciences*, *Business, Management and Accounting*, and *Economics, Econometrics and Finance*) for which we observed a decreasing trend in self-citations within the 2009-2016 publication window are different from those for which we observed a similar non-increasing trend in the 1957-2016 publication window: *Chemistry*, *Neuroscience*, and *Pharmacology, Toxicology, and Pharmaceutical Science*. This result may be since, according to our data, the guidelines imposed by the 2012 Italian Habilitation did not result in a substantial change in self-citations habits in the short term for the three disciplines mentioned above of the 2009-2016 period.

Along the same lines, it is worth mentioning that the discipline that had the second-best increment considering the 1957-2016 period, i.e. *Earth and Planetary Science*, had a smaller increment in the 2009-2016 period (from 4.74 a.sc.a. to 5.9 a.sc.a.) – which means that another decisive increasing trend for the discipline could have happened before 2009. We can envision a similar outcome also for the discipline *Energy*, which recorded an essential increment of self-citations according to the 1957-2016 citation data. The analysis of the 2009-2016 period did not entirely support this increment due to the controversial expected vs actual trend in author self-citations, as introduced in the previous section. Among the possible explanations for this situation, it is even possible that some disciplines started to adapt their self-citation behaviour to the new rules of the Italian Scientific Habilitation even before 2012. It is worth mentioning that scholars and policymakers discussed the Italian Law 240/2010 that introduced the new rules for the evaluation of academics in Italy for years before its formal adoption in 2011. Thus, it could be possible that some disciplines started to react to the first drafts of these new rules even before 2009.

A stronger conclusion can be reached, instead, on the ten disciplines that shown (in Fig. 4) a confidence interval greater than 0 and a non-controversial increasing author self-citations trend after 2012 (Fig. 5 and Fig. 6) in both the analysed period, i.e. 1957-2016 and 2009-2016. These disciplines were: *Arts and Humanities*, *Chemical Engineering*, *Decision Sciences*, *Engineering*, *Materials Science*, *Mathematics*, *Pharmacology, Toxicology and Pharmaceutical Science*, *Psychology*, *Social Sciences*, and *Veterinary Science and Veterinary Medicine*. In these cases, we think that there could exist, indeed, a causal relationship between the increment of author self-citations after 2012 and the rules introduced by the 2012 Italian Scientific Habilitation.

## Limitations of our study

The outcomes of our study provide additional insights into the debate about self-citations. However, we are aware of particular limitations that may have affected some of the interpretations we suggested, and that we list here to envision ways to improve our research in further studies.

The first limitation concerns the choice of the document set used for retrieving citation data. As explained in Section "Methods and Material", we have used all the bibliographic references of all the articles available on ScienceDirect in XML written by the participants of the 2012 Italian Scientific Habilitation. On the one hand, we think that this choice provided us with an excellent and representative population of articles and allowed us to have a corpus of documents we can use to further study citations in terms of their citation functions, for instance. On the other hand, it resulted in a limited number of documents. Consequently, we retrieved fewer citations than those included in the set of documents compliant with our selection criteria available in existing citation databases. These databases include: (a) commercial databases, such as Scopus and Web of Science, that we decided not to use due to limitations we had in using their APIs; and (b) open databases, such as Microsoft Academic Graph and Dimensions, that were not available in 2016 when we collected the data). The uses of these databases could help in different ways – from the availability of a more significant number of citations to the possibility of having authors already disambiguated, which is a crucial issue, in particular, for the identification of the author network self-citations.

Another issue is that we did not perform an international comparison about self-citations to confirm whether the observed increasing behaviour after 2012 is a peculiarity of the Italian system or a reflection of patterns that are working at an international level. Such an additional study would have, in principle, allowed us to strengthen our results. However, a recent work by Baccini et al. (2019) provides useful insights on this regard. They also observed a notable increasing trend in the inwardness (i.e. the proportion of citations coming from a country over the total number of citations gathered by that country) in Italy, comparing it with other ten countries within the same time window. Also, the self-citation increment seemed not to relate with an increase in the international collaboration, that remained quite stable in Italy in the years considered in the analysis, which overlap with the time window we used in our study. Indeed, since 2010, Italy has become "the European country with the lowest international collaboration and the highest inwardness" (Baccini et al. 2019).

In principle, other aspects could be influential for the outcomes of our analysis. For instance, an increase in publications by Italian scholars, thus following a publish-or-perish prescription, could justify an increase in self-citations – since the more publications one publishes, the more items one can self-cite. Similarly, an increase in the number of authors in publications could again justify an increase in self-citations. However, past studies such as (Aksnes 2003) showed that, even if the number of self-citations received by multi-authored articles is higher than that obtained by single-author articles, the overall citations received by such multi-authored articles is disproportionally high. It can be only partially justified by self-citations. We did not directly consider all these aspects in our analysis and, thus, they deserve to be studied further.

Finally, the last point we want to mention is the particular perspective used in our analysis. We preferred to study self-citations at the article level, by marking a citation between two articles as a self-citation when such articles shared a particular feature (either an author or an author network). We did not adopt an author-centric perspective, where we should have counted the self-citations performed by each author of any article in our corpus, and that would have resulted in an entirely different analysis. We think our approach was right for enabling the study of behavioural changes in self-citing at a global Italian level. However, it did not allow us to catch possible anomalous self-citing behaviours of authors – e.g. those who self-cite too much compared with the average of the other authors. Neither, it allowed us to filter the self-citations involving citing and cited articles authored only by the participants to the 2012 Italian Scientific Habilitation – or, more broadly, involving only scholars working in the Italian system. These aspects could be the object of further analysis.

## Conclusions

In this work, we have analysed citation data of the XML sources of all the articles available in ScienceDirect (co-)authored by the participants to the 2012 Italian Scientific Habilitation. All the data have been ingested, stored, and extended using Semantic Publishing technologies. We had analysed the increments/decrements and trends in author self-citations and author network self-citations before and after 2012, i.e. the first year when we adopted the new rules introduced by the Italian Government for assessing scholars. Our analysis showed an overall increment in author self-citations in several of the academic disciplines considered, but we observed a stronger causal relation between such increment and the rules introduced by the 2012 Italian Scientific Habilitation in 10 disciplines (RQ1). Instead, the analysis of the author network self-citations had a limited impact overall and was inconclusive for providing a clear answer to RQ2, mainly due to the lack of data for computing the co-author network of an author. We plan to explore this latter aspect further in future studies by using additional data coming from other open datasets (e.g. OpenCitations and Wikidata).

In principle, the outcomes of this study, as well as of similar studies which may take into consideration also other international settings, could be used to reduce the impact that self-citations can have when assessing a particular author or an entire discipline. For instance, one could use the mean of author self-citations of a particular discipline for coercing and, even, reducing the impact of the articles that show an exaggerated behaviour in such self-citations. A similar rationale could also be used for author network self-citations, to try to identify possible cliques that work to boost the citations of all their members. However, these approaches could also be affected by a strong bias, since they tend to punish the *amasser outliers*, i.e. those who self-cite too much, without rewarding the *shy outliers*, i.e. those who do not adopt such self-referential behaviours.

While the application of such average count of self-citation is a possibility, their systematic and indiscriminate use could sometimes be simplistic since it would not take into consideration situations where the use of a large number of self-citations is reasonable and justifiable in an article. Past works such as (Glänzel et al. 2006) suggest that self-citations should not be condemned by removing them from bibliometric statistics, even if they could distort and even falsifying the impact of a particular work when such measures are used to reward scholars. In order to control such distortion, we think that it could be more valuable to understand if these self-citations are either *organic*, *perfunctory* (Swales 1986) or even *coercive* citations (Yu, Yu & Wang 2014). Organic citations are those where the citing work needs the cited work for being fully understood. On the contrary, in the perfunctory/coercive citations, the cited work is mere acknowledgement done for some unclear (and even unspecified) reasons, and it is not functional to the understanding of the citing article. We do not have figures that show the mean of organic/perfunctory/coercive author self-citations in articles explicitly, even if we will consider this analysis for a future follow-up work. However, it is clear that articles that specify only organic author self-citations exist – for instance, when one describes her methods that are used to run some specific experiment, or an incremental study based on previous results. In addition to this discussion, past studies have also highlighted how self-citations, generally speaking, are not harmful and, instead, are actually a useful mechanism of knowledge diffusion (Gálvez 2017).

Along the lines above, informal colloquia we had with some of our colleagues in different disciplines seemed to show that a large part of self-citations is organic. Besides, external constraints (e.g. page limits and other editorial mandates) could drive the ratio between author self-citations and the total number of citations in an article. In these cases, if an author must choose to keep in the paper an organic self-citation or a perfunctory "normal" citation, it would be probable (and reasonable) to discard the normal one, thus keeping the self-citation.

Considering all these possible scenarios, we think it is unfair to systematically remove self-citations from the material used for addressing assessment exercises. Instead, we think a more sophisticated approach is needed when authors' reward is the final goal – such as to consider only organic citations in the computation of reward metrics. It is a complex issue that needs further studies and analysis to understand the actual feasibility clearly – e.g. can we develop an automatic approach to identify organic citations? – and efficacy – e.g. does the use of only organic citations contain the possible distortion effectively in the reward systems produced by considering all the citations to scholars' works?

# Appendix

| category | # p year <= 2012 | mean p year <= 2012 | std p year <= 2012 | # p year > 2012 | mean p year > 2012 | std p year > 2012 | diff | ci-low | ci-high |
|---|---|---|---|---|---|---|---|---|---|
| ALL | 26951 | 2.36 | 3.66 | 8722 | 3.18 | 5.33 | 0.82 | 0.72 | 0.92 |
| Agricultural and Biological Sciences | 1934 | 3.09 | 4.7 | 1031 | 3.33 | 5.9 | 0.23 | -0.16 | 0.62 |
| Arts and Humanities | 58 | 1.66 | 2.11 | 9 | 4.33 | 3.77 | 2.68 | 0.98 | 4.38 |
| Biochemistry, Genetics and Molecular Biology | 10974 | 2.65 | 3.97 | 2605 | 3.02 | 5.75 | 0.37 | 0.18 | 0.56 |
| Business, Management and Accounting | 78 | 1.69 | 3.08 | 77 | 1.91 | 2.87 | 0.22 | -0.73 | 1.16 |
| Chemical Engineering | 1582 | 1.94 | 3.03 | 864 | 2.33 | 3.74 | 0.39 | 0.12 | 0.67 |
| Chemistry | 4198 | 1.98 | 3.23 | 1021 | 1.9 | 3.93 | -0.07 | -0.3 | 0.16 |
| Computer Science | 785 | 1.83 | 2.59 | 326 | 2.69 | 4.13 | 0.86 | 0.46 | 1.26 |
| Decision Sciences | 348 | 2.54 | 2.91 | 147 | 4.16 | 4.01 | 1.62 | 0.99 | 2.25 |
| Earth and Planetary Sciences | 796 | 3.45 | 4.77 | 400 | 5.9 | 6.97 | 2.45 | 1.77 | 3.12 |
| Economics, Econometrics and Finance | 135 | 0.82 | 1.29 | 28 | 1.21 | 1.97 | 0.39 | -0.19 | 0.98 |
| Energy | 668 | 1.02 | 1.8 | 783 | 2.61 | 3.61 | 1.59 | 1.28 | 1.89 |
| Engineering | 1770 | 1.57 | 2.43 | 1204 | 3.01 | 3.94 | 1.44 | 1.21 | 1.67 |
| Environmental Science | 1699 | 2.1 | 3.42 | 800 | 2.86 | 4.65 | 0.76 | 0.43 | 1.08 |
| Immunology and Microbiology | 1704 | 3.03 | 3.93 | 661 | 3.28 | 7.42 | 0.24 | -0.22 | 0.71 |
| Materials Science | 1463 | 2.44 | 3.11 | 484 | 2.45 | 3.99 | 0.02 | -0.33 | 0.36 |
| Mathematics | 1401 | 1.66 | 2.25 | 390 | 2.92 | 3.97 | 1.26 | 0.96 | 1.57 |
| Medicine and Dentistry | 10147 | 1.98 | 3.45 | 3434 | 2.81 | 5.48 | 0.83 | 0.68 | 0.99 |
| Neuroscience | 2723 | 2.47 | 3.58 | 768 | 2.25 | 4.23 | -0.21 | -0.51 | 0.09 |
| Nursing and Health Professions | 613 | 2.74 | 3.34 | 117 | 3.5 | 5.92 | 0.75 | -0.01 | 1.52 |
| Pharmacology, Toxicology and Pharmaceutical Science | 2586 | 1.74 | 3.3 | 538 | 1.56 | 4.65 | -0.18 | -0.51 | 0.16 |
| Physics and Astronomy | 1185 | 2.45 | 3.48 | 346 | 4.07 | 4.68 | 1.63 | 1.17 | 2.08 |
| Psychology | 552 | 2.76 | 3.25 | 294 | 3.88 | 4.94 | 1.13 | 0.57 | 1.68 |
| Social Sciences | 335 | 2.2 | 3.45 | 210 | 2.88 | 4.39 | 0.68 | 0.01 | 1.34 |
| Veterinary Science and Veterinary Medicine | 341 | 3.21 | 3.82 | 135 | 5.48 | 6.17 | 2.27 | 1.35 | 3.19 |

*Table 1* The citing articles considered in the experiment (published between 1957 and 2016), split into two populations and by subject category: articles published by 2012 [year <= 2012], and articles published after 2012 [year > 2012]. In addition to the number of the articles included in the two populations, the table shows the mean of the author self-citations per article, the related standard deviation, the difference of the means between the two populations accompanied by the related confidence interval "ci-low" / "ci-high".

| category | # p year <= 2012 | mean p year <= 2012 | std p year <= 2012 | # p year > 2012 | mean p year > 2012 | std p year > 2012 | diff | ci-low | ci-high |
|---|---|---|---|---|---|---|---|---|---|
| ALL | 8511 | 2.45 | 4.37 | 8722 | 3.18 | 5.33 | 0.74 | 0.59 | 0.88 |
| Agricultural and Biological Sciences | 750 | 3.59 | 5.44 | 1031 | 3.33 | 5.9 | -0.26 | -0.8 | 0.27 |
| Arts and Humanities | 22 | 1.77 | 2.13 | 9 | 4.33 | 3.77 | 2.56 | 0.39 | 4.73 |
| Biochemistry, Genetics and Molecular Biology | 2732 | 2.54 | 5.17 | 2605 | 3.02 | 5.75 | 0.47 | 0.18 | 0.77 |
| Business, Management and Accounting | 39 | 2.13 | 3.93 | 77 | 1.91 | 2.87 | -0.22 | -1.49 | 1.05 |
| Chemical Engineering | 762 | 1.75 | 3.01 | 864 | 2.33 | 3.74 | 0.58 | 0.25 | 0.91 |
| Chemistry | 1102 | 1.11 | 2.87 | 1021 | 1.9 | 3.93 | 0.79 | 0.5 | 1.08 |
| Computer Science | 278 | 2 | 3.01 | 326 | 2.69 | 4.13 | 0.7 | 0.11 | 1.28 |
| Decision Sciences | 134 | 3.44 | 3.75 | 147 | 4.16 | 4.01 | 0.72 | -0.2 | 1.63 |
| Earth and Planetary Sciences | 326 | 4.74 | 5.91 | 400 | 5.9 | 6.97 | 1.16 | 0.2 | 2.11 |
| Economics, Econometrics and Finance | 33 | 1.24 | 1.65 | 28 | 1.21 | 1.97 | -0.03 | -0.96 | 0.9 |
| Energy | 464 | 1.04 | 1.78 | 783 | 2.61 | 3.61 | 1.57 | 1.22 | 1.92 |
| Engineering | 865 | 1.9 | 2.88 | 1204 | 3.01 | 3.94 | 1.11 | 0.81 | 1.42 |
| Environmental Science | 717 | 2.32 | 4.12 | 800 | 2.86 | 4.65 | 0.54 | 0.09 | 0.98 |
| Immunology and Microbiology | 553 | 2.42 | 4.06 | 661 | 3.28 | 7.42 | 0.85 | 0.16 | 1.55 |
| Materials Science | 381 | 1.49 | 2.67 | 484 | 2.45 | 3.99 | 0.96 | 0.49 | 1.43 |
| Mathematics | 409 | 1.91 | 2.68 | 390 | 2.92 | 3.97 | 1 | 0.54 | 1.47 |
| Medicine and Dentistry | 3666 | 2.16 | 4.13 | 3434 | 2.81 | 5.48 | 0.65 | 0.42 | 0.87 |
| Neuroscience | 770 | 2.07 | 3.92 | 768 | 2.25 | 4.23 | 0.18 | -0.23 | 0.59 |
| Nursing and Health Professions | 123 | 3.26 | 4.8 | 117 | 3.5 | 5.92 | 0.24 | -1.13 | 1.6 |
| Pharmacology, Toxicology and Pharmaceutical Science | 714 | 1.09 | 3.63 | 538 | 1.56 | 4.65 | 0.48 | 0.02 | 0.93 |
| Physics and Astronomy | 277 | 2.79 | 4.63 | 346 | 4.07 | 4.68 | 1.29 | 0.55 | 2.02 |
| Psychology | 224 | 3.01 | 3.64 | 294 | 3.88 | 4.94 | 0.87 | 0.1 | 1.64 |
| Social Sciences | 158 | 2.55 | 4.22 | 210 | 2.88 | 4.39 | 0.33 | -0.56 | 1.22 |
| Veterinary Science and Veterinary Medicine | 140 | 3.99 | 4.49 | 135 | 5.48 | 6.17 | 1.49 | 0.21 | 2.77 |

***Table 2*** *The citing articles considered in the experiment (published between 2009 and 2016), split into two populations and by subject category: articles published by 2012 [year <= 2012], and articles published after 2012 [year > 2012]. In addition to the number of the articles included in the two populations, the table shows the mean of the author self-citations per article, the related standard deviation, the difference of the means between the two populations accompanied by the related confidence interval "ci-low" / "ci-high".*

| category | # p year <= 2012 | mean p year <= 2012 | std p year <= 2012 | # p year > 2012 | mean p year > 2012 | std p year > 2012 | diff | ci-low | ci-high |
|---|---|---|---|---|---|---|---|---|---|
| ALL | 26951 | 0.54 | 2.22 | 8722 | 0.77 | 3.06 | 0.23 | 0.17 | 0.29 |
| Agricultural and Biological Sciences | 1934 | 0.86 | 3.34 | 1031 | 0.85 | 4.43 | -0.01 | -0.29 | 0.28 |
| Arts and Humanities | 58 | 0.55 | 1.28 | 9 | 0.56 | 0.96 | 0 | -0.88 | 0.89 |
| Biochemistry, Genetics and Molecular Biology | 10974 | 0.56 | 2.38 | 2605 | 0.81 | 3.21 | 0.25 | 0.14 | 0.35 |
| Business, Management and Accounting | 78 | 0.08 | 0.47 | 77 | 0.18 | 0.92 | 0.11 | -0.13 | 0.34 |
| Chemical Engineering | 1582 | 0.34 | 1.12 | 864 | 0.41 | 1.5 | 0.07 | -0.03 | 0.18 |
| Chemistry | 4198 | 0.25 | 1.28 | 1021 | 0.29 | 1.32 | 0.03 | -0.05 | 0.12 |
| Computer Science | 785 | 0.29 | 1.24 | 326 | 0.7 | 2.65 | 0.4 | 0.17 | 0.63 |
| Decision Sciences | 348 | 0.24 | 0.76 | 147 | 0.33 | 0.98 | 0.09 | -0.07 | 0.25 |
| Earth and Planetary Sciences | 796 | 1.2 | 3.4 | 400 | 1.7 | 6.33 | 0.5 | -0.06 | 1.05 |
| Economics, Econometrics and Finance | 135 | 0.02 | 0.17 | 28 | 0.32 | 1.17 | 0.31 | 0.1 | 0.51 |
| Energy | 668 | 0.2 | 0.73 | 783 | 0.37 | 1.2 | 0.17 | 0.07 | 0.28 |
| Engineering | 1770 | 0.18 | 0.76 | 1204 | 0.41 | 1.77 | 0.22 | 0.13 | 0.32 |
| Environmental Science | 1699 | 0.44 | 1.67 | 800 | 0.63 | 2.15 | 0.19 | 0.04 | 0.35 |
| Immunology and Microbiology | 1704 | 0.69 | 2.14 | 661 | 1.16 | 4.88 | 0.46 | 0.18 | 0.75 |
| Materials Science | 1463 | 0.23 | 0.86 | 484 | 0.42 | 1.62 | 0.19 | 0.07 | 0.3 |
| Mathematics | 1401 | 0.18 | 0.79 | 390 | 0.86 | 3.37 | 0.69 | 0.49 | 0.88 |
| Medicine and Dentistry | 10147 | 0.54 | 1.93 | 3434 | 0.79 | 2.97 | 0.25 | 0.16 | 0.34 |
| Neuroscience | 2723 | 0.56 | 1.91 | 768 | 0.51 | 2.25 | -0.05 | -0.2 | 0.11 |
| Nursing and Health Professions | 613 | 0.84 | 2.12 | 117 | 0.69 | 2.23 | -0.15 | -0.57 | 0.28 |
| Pharmacology, Toxicology and Pharmaceutical Science | 2586 | 0.26 | 1.39 | 538 | 0.37 | 2.51 | 0.11 | -0.05 | 0.26 |
| Physics and Astronomy | 1185 | 0.53 | 4.1 | 346 | 0.69 | 2.11 | 0.16 | -0.29 | 0.61 |
| Psychology | 552 | 0.72 | 1.84 | 294 | 0.67 | 1.85 | -0.05 | -0.31 | 0.22 |
| Social Sciences | 335 | 0.65 | 1.95 | 210 | 0.39 | 1.81 | -0.27 | -0.6 | 0.06 |
| Veterinary Science and Veterinary Medicine | 341 | 0.74 | 1.67 | 135 | 0.84 | 2.1 | 0.11 | -0.25 | 0.47 |

***Table 3*** *The citing articles considered in the experiment (published between 1957 and 2016), split into two populations and by subject category: articles published by 2012 [year <= 2012], and articles published after 2012 [year > 2012]. In addition to the number of the articles included in the two populations, the table shows the mean of the author network self-citations per article, the related standard deviation, the difference of the means between the two populations accompanied by the related confidence interval "ci-low" / "ci-high".*

| category | # p year <= 2012 | mean p year <= 2012 | std p year <= 2012 | # p year > 2012 | mean p year > 2012 | std p year > 2012 | diff | ci-low | ci-high |
|---|---|---|---|---|---|---|---|---|---|
| ALL | 8511 | 0.77 | 3.1 | 8722 | 0.77 | 3.06 | 0.01 | -0.09 | 0.1 |
| Agricultural and Biological Sciences | 750 | 1.27 | 4.22 | 1031 | 0.85 | 4.43 | -0.42 | -0.83 | -0.01 |
| Arts and Humanities | 22 | 0.96 | 1.43 | 9 | 0.56 | 0.96 | -0.4 | -1.46 | 0.67 |
| Biochemistry, Genetics and Molecular Biology | 2732 | 0.86 | 3.66 | 2605 | 0.81 | 3.21 | -0.05 | -0.24 | 0.13 |
| Business, Management and Accounting | 39 | 0.15 | 0.66 | 77 | 0.18 | 0.92 | 0.03 | -0.3 | 0.36 |
| Chemical Engineering | 762 | 0.36 | 1.08 | 864 | 0.41 | 1.5 | 0.06 | -0.07 | 0.19 |
| Chemistry | 1102 | 0.29 | 1.6 | 1021 | 0.29 | 1.32 | 0 | -0.13 | 0.13 |
| Computer Science | 278 | 0.38 | 1.3 | 326 | 0.7 | 2.65 | 0.32 | -0.02 | 0.66 |
| Decision Sciences | 134 | 0.44 | 1.04 | 147 | 0.33 | 0.98 | -0.11 | -0.34 | 0.13 |
| Earth and Planetary Sciences | 326 | 2.06 | 4.69 | 400 | 1.7 | 6.33 | -0.36 | -1.19 | 0.47 |
| Economics, Econometrics and Finance | 33 | 0.06 | 0.34 | 28 | 0.32 | 1.17 | 0.26 | -0.17 | 0.69 |
| Energy | 464 | 0.19 | 0.68 | 783 | 0.37 | 1.2 | 0.17 | 0.05 | 0.29 |
| Engineering | 865 | 0.25 | 0.89 | 1204 | 0.41 | 1.77 | 0.15 | 0.03 | 0.28 |
| Environmental Science | 717 | 0.63 | 2.21 | 800 | 0.63 | 2.15 | 0 | -0.22 | 0.22 |
| Immunology and Microbiology | 553 | 0.91 | 2.86 | 661 | 1.16 | 4.88 | 0.25 | -0.22 | 0.71 |
| Materials Science | 381 | 0.27 | 0.85 | 484 | 0.42 | 1.62 | 0.14 | -0.04 | 0.32 |
| Mathematics | 409 | 0.32 | 1.03 | 390 | 0.86 | 3.37 | 0.54 | 0.2 | 0.89 |
| Medicine and Dentistry | 3666 | 0.7 | 2.34 | 3434 | 0.79 | 2.97 | 0.1 | -0.03 | 0.22 |
| Neuroscience | 770 | 0.54 | 1.66 | 768 | 0.51 | 2.25 | -0.03 | -0.23 | 0.17 |
| Nursing and Health Professions | 123 | 1.12 | 2.71 | 117 | 0.69 | 2.23 | -0.43 | -1.06 | 0.2 |
| Pharmacology, Toxicology and Pharmaceutical Science | 714 | 0.31 | 1.58 | 538 | 0.37 | 2.51 | 0.06 | -0.17 | 0.29 |
| Physics and Astronomy | 277 | 1.08 | 7.85 | 346 | 0.69 | 2.11 | -0.4 | -1.26 | 0.47 |
| Psychology | 224 | 0.8 | 1.89 | 294 | 0.67 | 1.85 | -0.13 | -0.45 | 0.2 |
| Social Sciences | 158 | 0.82 | 2.15 | 210 | 0.39 | 1.81 | -0.44 | -0.84 | -0.03 |
| Veterinary Science and Veterinary Medicine | 140 | 1.01 | 1.97 | 135 | 0.84 | 2.1 | -0.17 | -0.65 | 0.31 |

***Table 4*** *The citing articles considered in the experiment (published between 2009 and 2016), split into two populations and by subject category: articles published by 2012 [year <= 2012], and articles published after 2012 [year > 2012]. In addition to the number of the articles included in the two populations, the table shows the mean of the author network self-citations per article, the related standard deviation, the difference of the means between the two populations accompanied by the related confidence interval "ci-low" / "ci-high".*